# A Stern-Gerlach experiment for slow light

Leon Karpa[1] and Martin Weitz[1,2,3]

[1]*Physikalisches Institut der Universität Tübingen, Auf der Morgenstelle 14, 72076 Tübingen, Germany*

[2]*Center of Ultracold Atoms, Massachusetts Institute of Technology, Cambridge, MA 02139, USA*

[3]*Institut für Angewandte Physik der Universität Bonn, Wegelerstrasse 8, 53115 Bonn, Germany*

**Electromagnetically induced transparency allows for light transmission through dense atomic media by means of quantum interference[1]. Media exhibiting electromagnetically induced transparency have very interesting properties, such as extremely slow group velocities[2-4]. Associated with the slow light propagation are quasiparticles, the so-called dark polaritons, which are mixtures of a photonic and an atomic contribution[5]. We here demonstrate that these excitations behave as particles with a nonzero magnetic moment, which is in clear contrast to the properties of a free photon. It is found that light passing through a rubidium gas cell under the conditions of electromagnetically induced transparency is deflected by a small magnetic field gradient. The deflection angle is proportional to the optical propagation time through the cell. The observed beam deflection can be understood by assuming that dark state polaritons have an effective magnetic moment. Our experiment can be described in terms of a Stern-Gerlach experiment for the polaritons.**

The magnetic dipole moment of a particle is a fundamental property. Its magnitude is different from zero for all known massive elementary particles. In contrast, no magnetic moment exists for photons in vacuum[6]. It is the purpose of this letter to demonstrate that when light propagates though matter it can behave as if it has acquired a magnetic momentum. Notably, in the field of solid state physics it is well known that the environment can modify a particle's properties, and common concepts here include e.g. the effective mass or charge of an electron[7].

In our experiment, we study the deflection of a circularly polarized optical beam in a Stern Gerlach-like magnetic field gradient under the conditions of electromagnetically induced transparency in a rubidium vapor cell. We find that the optical beam is deflected by an angle of typically several $10^{-5}$ rad, with a smaller group velocity leading to a larger deflection. We interpret our results in terms of the dark polariton, as the quasiparticle that is physically associated with the slow light propagation, having a nonzero magnetic moment. The deflection can be understood in a simple model by considering the mechanical force $\mathbf{F} = \nabla\,(\boldsymbol{\mu}_{pol} \cdot \mathbf{B}(\mathbf{r}))$ onto the hybrid light-matter quasiparticle with magnetic moment $\boldsymbol{\mu}_{pol}$ in the spatially inhomogeneous Stern-Gerlach magnetic field $\mathbf{B}(\mathbf{r})$. The origin of the magnetic moment is attributed to the spin wave contribution of the polariton that develops upon the entry of light into the medium, and in agreement with theoretical expectations.

Before proceeding, we note that it is well known that the spectral position of dark resonances is very sensitive to magnetic fields, and both absorptive and dispersive properties of these resonances can be used for magnetometry[8]. Slow light experiments directly illustrate the highly dispersive nature of electromagnetically induced transparency[2-4]. More recently, a stopping of light was also demonstrated[9,10]. We furthermore wish to point out that the deflection of optical beams in external fields has been subject to earlier works[11-13]. In recent experiments, angle deflection was observed caused by spatially inhomogeneous optical pumping and saturation effects[12,13]. In other work, beam focusing has been achieved by a spatially inhomogeneous modification of the refractive index with a pump beam[14]. In contrast to all earlier works, the effect observed here gives access to the intrinsic magnetic dipole moment of a light-matter quasiparticle. The obtained optical beam deflection is interpreted in terms of the Stern-Gerlach effect for the quasiparticles.

Our experiment is carried out in a heated buffer gas cell with light tuned to the rubidium ($^{87}$Rb) D1-line. The observed effects can be understood by considering the Λ-type three-level configuration shown in Fig. 1a. The levels are coupled by two optical fields, a weaker "signal" field and a stronger "control" field. The beams have opposite circular polarisations, and couple the two stable ground states |g-> and |g+>, with Zeeman quantum numbers differing by two, over a short lived electronically excited state |e>. When the atomic population is left in a suitable ('dark') coherent superposition of the ground states, the excitation amplitudes into the upper state interfere destructively and the atomic medium is transparent to light[1]. Of importance to our experiment is that spontaneous processes can thus (in principle) be completely avoided, and coherence be preserved. It is well known that besides spectrally sharp variations in absorption (note that a truly stationary dark state only exists for a two-photon detuning δ of zero), also drastic modifications in the dispersive properties occur, which are the origin of slow light propagation.

In our experiment, the observed optical group velocities are typically a few hundreds m/s, which results in a Stern-Gerlach deflection of the light-matter quasiparticles some 12 orders of magnitude larger compared to a situation in which such a quasiparticle would move with a speed close to vacuum light velocity. A transverse magnetic field gradient, designed to be compatible to the presence of electromagnetically induced transparency, is applied to the rubidium cell (see methods). As shown in Fig. 1b, the bias field is directed collinearly to the signal and control beams (the latter is not shown in this figure), which we have denoted as the z-axis in this figure. The magnitude of the bias field varies in a direction transverse to the beam, and was denoted as the x-axis. For magnetic dipoles oriented along the beam axis a non-zero transverse force is expected. After its passage though the cell, we monitor the transverse displacement of the signal beam.

In initial measurements, we have mapped out the magnetic field by recording dark resonances for variable transverse positions of the beam. A typical spectrum is shown in Fig. 2a as a function of the two-photon detuning. The field gradient was found to be relatively uniform along the transverse direction. We are aware that our measurement technique is not very sensitive to residual longitudinal gradients.

Subsequently, we have monitored the angle deflection of the signal beam while scanning though a dark resonance. For this measurement, the beam position was monitored with a CCD-camera placed 2 m apart from the cell. Fig. 2b shows the result of a corresponding experiment, in which the beam on two-photon resonance is deflected by 35 μm, corresponding to an angle deflection near $2 \cdot 10^{-5}$ rad. The observed spectrally sharp deflection peak is accompanied by two side maxima with opposite deflection direction. With other measurements, verifying both the absence of a deflection with no field gradient and monitoring the beam position in a different focal plane, we have confirmed that indeed field gradients give rise to a beam angle deflection, as is expected for a non-zero magnetic moment of the dark state polariton. This reaches its maximum absolute value near zero two-photon detuning, where also the transparency is maximal. We interpret our results in terms of a proof of principle experiment demonstrating that the dark polariton exhibits an effective magnetic moment. The presence of such a dipole moment is attributed to the atomic spin-wave contribution.

In a mechanical model, the expected angle deflection of a polariton subject to a transverse magnetic force $\frac{dB_z}{dx}$ for a dipole moment $\mu_{pol}$ aligned collinearly to the bias field is for small deflection angles given by $\alpha \cong F\, t_{int}/p$ with $F = \frac{dB_z}{dx} \mu_{pol}$. Here, $t_{int} = L/v_g$ denotes the interaction time with L as the cell length ($\cong 50$ mm here), $v_g$ the optical group velocity and p = nhk the photon momentum[15], for which we take the free space value since the refractive index n here is very close to unity. We arrive at an expected deflection angle

$$\alpha \cong \frac{L}{v_g} \cdot \frac{\mu_{pol}}{p} \cdot \frac{dB_z}{dx}, \qquad (1)$$

which is proportional to the time that a dark polariton is subject to the deflecting field gradient. To monitor the experimental dependence of the deflection on this interaction time, we have recorded the beam deflection at two-photon resonance for different control laser intensities as to vary the optical group velocity[5]. In a separate measurement, for each of those control laser intensities the corresponding optical group velocity was measured by observing the delay that a gaussian-shaped signal beam pulse experienced when passing though the cell. Fig. 3 now shows the observed beam angle deflection as a function of the inverse group velocity. Here, an increase of the deflection is clearly observed for larger values of $1/v_g$, corresponding to longer interaction times with the field gradient. On the other hand, the experimental data does not show the linear dependence predicted by our simple model. We attribute this to an inhomogeneous variation of the two-photon detuning inherent both in our Stern-Gerlach deflection measurement due to the finite optical beam diameter transversally to the field gradient and our pulsed group velocity measurement due to the Fourier width of the pulses. This causes some loss of coherence of the fragile dark state and nonlinearities of the dispersion. From our observed Stern-Gerlach deflection, our simple model nevertheless allows us to extract an estimate for the value of the unperturbed polariton magnetic moment by solving eq. 1 for this quantity. From the average of our deflection data at a magnetic field gradient of $\frac{dB_z}{dx} \cong 910$ µG/mm (this gradient was achieved at a magnetic field on the beam axis of 116 mG), we estimate a value of $5.1 \cdot 10^{-24}$ J/T for the magnetic moment. The expected value for this moment can be calculated both directly from the dark polariton model and alternatively indirectly via the deflection originating from the transversally dependent dispersive properties of the dark resonance, and in the limit $v_g \ll c$ is given by one Bohr magnetron for the used atomic transition (see methods). Note that also for a 'usual' Stern-Gerlach experiment carried out e.g. with a neutron beam the observed angle deflection can also be described alternatively in particle or wave-nature pictures (the latter here certainly being less common), as follows from the well known quantum-mechanical wave-particle duality. Our experimental value for the dark polariton magnetic moment is a factor of 1.8 lower than the theoretical result, and thus clearly of correct order of magnitude. For the future, it is believed that a refined theoretical model accounting for the finite control beam intensity and connecting the effective polariton interaction time with our pulsed group velocity measurement is required. Ideally, the experiments should be carried out with an ultracold rubidium source, which should allow for very detailed comparisons of theory and experiment.

We expect that the observed effects can have applications in the field of quantum information storage and manipulation in optical gases for switching applications and the addressing of parallel channels[16-18]. A different perspective would be to search for signatures of the Aharonov-Casher effect with slow light originating from the interaction of the effective magnetic dipole moment with an external electric field[19,20]. Such an experiment could further elucidate the quantum properties of the dark polariton.

**Methods**

**Experimental setup**

Our setup is a modified version of a previously described apparatus[21]. Both signal and control optical beams are derived from the same grating-stabilized diode laser source operating near 795 nm. The two beams pass independent acousto-optical modulators to allow for a variation of the optical difference frequency, are spatially overlapped and then fed through an optical fiber. Subsequently, the collinear beams are expanded to a 2mm beam diameter, and with oppositely circular polarizations sent through a rubidium cell subject to the Stern-Gerlach field. We use a heated rubidium gas cell filled with 20 torr Ne buffer gas. To generate the desired magnetic gradient field, we place a µ-metal stripe mounted parallel to the optical beams into an otherwise homogeneous magnetic field. This ferromagnetic stripe is considerably longer than the cell, and acts as a shortcut for the magnetic flux. At the cell side directed to this stripe, the field is lowered which results in a transverse magnetic field gradient. In our experiment, the control field intensity was usually much stronger than that of the signal field, so that most of the relevant population was in the F=2, $m_F$=-2 ground state sublevel. Under these circumstances, the states |g-⟩, |g+⟩ and |e⟩ of the simplified three-level system (Fig. 1a) correspond to the $m_F$=-2, F=2 and $m_F$=0, F=2 ground state and the $m_F$=1, F'=1 excited state sublevels of the rubidium D1-line components respectively.

**Theoretical model**

The deflection of the slow light beam can be described theoretically both by a wave-optics and a particle model respectively, with the latter incorporating the concept of the dark polariton. Let us begin this section by outlining the wave-optics description, for which we assume that the response of the ensemble of three-level atoms to the signal field is such that it can be described by a refractive index for the signal field. Transversely to the beam axis, the magnetic field gradient causes a spatial variation of the two-photon detuning of $2\, g_F\, \mu_B\, \frac{dB_z}{dx}$, where $\mu_B$ denotes the Bohr magnetron and $g_F$ the (hyperfine) g-factor, being equal to 1/2 for the used rubidium transition. In this model, the corresponding variation of the refractive index, determined by the magnitude of $\frac{dn}{d\omega}$, causes a prism-like angle deflection of the optical beam. One readily finds that the lineshape of the angle deflection thus should be proportional to the derivative of a dispersion shaped line, which qualitatively agrees well with the spectrum of Fig. 2b. Further, when using the known formula for the group velocity $v_g = c/(n+\omega \cdot \frac{dn}{d\omega})$, we for $v_g \ll c$ arrive at $\alpha \cong \frac{dB_z}{dx} \cdot 2\, g_F\, \mu_B \cdot \lambda\, L/(h\, v_g)$. Note that this result is identical to eq. 1 if we set $\mu_{pol} \cong 2\, g_F\, \mu_B$.

The observed optical beam deflection must also be present in a quantum-mechanical particle description, where the light propagation though the medium is described by the propagation of hybrid atom-light quasiparticles, i.e. dark polaritons for the EIT medium. To allow for a determination of the Stern-Gerlach force, the expected polariton magnetic moment is determined as follows. Consider a quantum signal field connecting the states $|g_-\rangle$ and $|e\rangle$ with coupling constant g and a classical control field driving the transition between levels $|g_+\rangle$ and $|e\rangle$ with a Rabi frequency $\Omega$ for an ensemble of N identical three-level atoms (Fig. 1a). Following [5,22], we define a mixing angle of atomic and photonic contributions via: $\tan \Theta = g \sqrt{N}/\Omega$, which is directly related to the optical group velocity: $v_g = c \cos^2 \Theta$. Starting from a polariton "vacuum state" with no such excitations: $|0\rangle_p = |0\rangle_{elm} |g_-...g_-\rangle$, where $|n\rangle_{elm}$ denotes a state with n photons in the signal mode and $|g_-...g_-\rangle$ a collective atomic state with all N atoms in the internal state $|g_-\rangle$, we can generate a one-polariton state $|1\rangle_p = \Psi^\dagger |0\rangle_p$ using the operator

$$\Psi^\dagger = \cos(\Theta)\, a^\dagger - \sin(\Theta) \frac{1}{\sqrt{N}} \sum_{j=1}^{N} \sigma_{-+}^j \ .$$

Here, $a^\dagger$ denotes the creation operator for a signal photon and $\sigma_{-+}^j$ the spin operator required to flip the j-th atom into state $|g_+\rangle$. In a calculation pointed out to us by M. Fleischhauer, the magnetic moment of a polariton can now be derived by considering the difference of the spin expectation values $S_z$ of a one-polariton state and the polariton vacuum. For a $\Delta m=2$ Raman transition, one arrives at $\mu_{pol}=2g_F\, \mu_B\, (_p\langle 1| S_z |1\rangle_p - _p\langle 0|$

$S_z |0>_p) = 2g_F \mu_B \sin^2 \Theta$. For slow light ($v_g \ll c$), this reduces to $\mu_{pol} = 2 g_F \mu_B$, which equals the value obtained indirectly from the above described classical model.

Finally, we give the corresponding values for the gyromagnetic ratio and the g-factor of the dark polariton. Since the angular moment of this quasiparticle is h, being equal to the photon spin, we readily find $\gamma = -2g_F\mu_B \sin^2 (\Theta) /h$ and $g_{pol} = 2 g_F \sin^2 \Theta$, respectively.

[1]    See e.g.: E. Arimondo, "Coherent population trapping in laser spectroscopy", Prog. Opt. **35**, 257 - 354 (1996).

[2]    L. V. Hau, S. E. Harris, Z. Dutton, and C. H. Behroozi, "Light speed reduction to 17 metres per second in an ultracold atomic gas", Nature (London) **397**, 594 - 598 (1999).

[3]    M. Kash et al., "Ultraslow group velocity and enhanced nonlinear optical effects in a coherently driven hot atomic gas", Phys. Rev. Lett. **82**, 5229 - 5232 (1999).

[4]    D. Budker, D. F. Kimball, S. M. Rochester, and V. V. Yashchuk, "Nonlinear Magneto-optics and Reduced Group Velocity of Light in Atomic Vapor with Slow Ground State Relaxation", Phys. Rev. Lett. **83**, 1767 - 1770 (1999).

[5]    M. Fleischhauer and M. D. Lukin, "Dark-state polaritons in electromagnetically induced transparency", Phys. Rev. Lett. **84**, 5094 - 5097 (2000).

[6]    See, e.g.: J. D. Jackson, "Classical electrodynamics", (Wiley, New York, 1975).

[7]    See, e.g.: N. W. Ashcroft and N. D. Mermin, "Solid state physics", (Saunders College Publishing, New York, 1976).

[8]    M. O. Scully and M. Fleischhauer, "High-sensitivity magnetometer based on index-enhanced media", Phys. Rev. Lett. **69**, 1360 - 1363 (1992).

[9]     C. Liu, Z. Dutton, C. H. Behroozi, and L. V. Hau, "Observation of coherent optical information storage in an atomic medium using halted light pulses", Nature **409**, 490 - 493 (2001).

[10]     D. F. Phillips, A. Fleischhauer, A. Mair, R. L. Walsworth, and M. D. Lukin, "Storage of Light in Atomic Vapor", Phys. Rev. Lett. **86**, 783 - 786 (2001).

[11]     R. Schlesser and A. Weis, "Light-beam deflection by cesium vapor in a transverse-magnetic field", Opt. Lett. **17**, 1015 - 1017 (1992).

[12]     R. Holzner et al., "Observation of Magnetic-Field-Induced Laser Beam Deflection in Sodium Vapor", Phys. Rev. Lett. **78**, 3451 - 3454 (1997).

[13]     G. T. Purves, G. Jundt, C. S. Adams, and I. G. Hughes, "Refractive index measurements by probe-beam deflection", Eur Phys J D **29**, 433 - 436 (2004).

[14]     R. R. Moseley, S. Shepherd, D. J. Fulton, B. D. Sinclair, and M. H. Dunn, "Spatial Consequences of Electromagnetically Induced Transparency: Observation of Electromagnetically Induced Focusing", Phys. Rev. Lett. **74**, 670 - 673 (1995).

[15]     G. K. Campbell et al., "Photon Recoil Momentum in Dispersive Media", Phys. Rev. Lett. **94**, 170403 (2005)

[16]     M. D. Lukin, S. F. Yelin, and M. Fleischhauer, "Entanglement of atomic ensembles by trapping correlated photon states", Phys. Rev. Lett. **84**, 4232 - 4235 (2000).

[17]     T. Chanelière, D. N. Matsukevich, S. D. Jenkins, S.-Y. Lan, T. A. B. Kennedy and A. Kuzmich, "Storage and retrieval of single photons transmitted between remote quantum memories", Nature **438**, 833 - 836 (2005).

[18]     M. D. Eisaman, A. André, F. Massou, M. Fleischhauer, A. S. Zibrov and M. D. Lukin, "Electromagnetically induced transparency with tunable single-photon pulses", Nature **438**, 837 - 841 (2005).

[19]    Y. Aharonov and A. Casher, "Topological Quantum Effects for Neutral Particles", Phys. Rev. Lett. **53**, 319 - 321 (1984).

[20]    See also: U. Leonhardt and P. Piwnick, "Relativistic Effects of Light in Moving Media with Extremely Low Group Velocity", Phys. Rev. Lett. **84**, 822 - 825 (2000).

[21]    C. Bolkart, D. Rostohar, and M. Weitz, "Dark resonances with variable Doppler sensitivity", Phys. Rev. A **71**, 043816 (2005).

[22]    M. Fleischhauer and M. D. Lukin, "Quantum memory for photons: Dark-state polaritons", Phys. Rev. A **65**, 022314 (2002).

**Supplementary Information** accompanies the paper on **www.nature.com/nphys**.'

**Acknowledgements** We are indebted to M. Fleischhauer for his direct calculation of the magnetic dipole moment from the dark polariton model and for discussions. M.W. acknowledges hospitality by the CUA during his guest stay at MIT. We acknowledge financial support from the Deutsche Forschungsgemeinschaft, the Landesstiftung Baden-Württemberg and the European Community.

**Competing interests statement** The authors declare that they have no competing financial interests.

**Correspondence** and requests for materials should be addressed to M. W. (e-mail: martin.weitz@uni-tuebingen.de).

**Figure 1** Principle of experiment. **a** Simplified diagram of relevant atomic levels coupled to a $\sigma^-$-polarized control field and a $\sigma^+$-polarized signal field. **b** Scheme of the experiment, in which a transverse Stern-Gerlach field gradient causes an angle deflection of the optical signal beam within a rubidium gas cell. This deflection is interpreted as to yield evidence for the dark polariton, as the relevant hybrid atom-light quasiparticle, to possess an effective

magnetic moment. The optical control beam (not shown in the figure) is, before entering the cell, aligned collinearly to the signal beam.

**Figure 2** Experimental spectra. **a** Signal beam intensity transmitted though the rubidium cell as a function of the two-photon detuning $\delta = \omega_s - \omega_c + \mu_B \cdot B$ from the Raman resonance, where $\omega_s$ ($\omega_c$) denotes the optical frequency of the signal (control) beam, $\mu_B$ the Bohr magnetron, and B the magnetic field on the beam axis. **b** Measured transverse position of the signal beam along the x-axis as a function of the two-photon detuning. On two-photon resonance, a deflection of the signal beam corresponding to a deflection angle near $2 \cdot 10^{-5}$ radians is observed.

**Figure 3** Deflection of the signal beam on two-photon resonance as a function of the inverse optical group velocity. The shown error bars for the beam deflection give the standard deviation of the mean value of 14 measurements for each point respectively. The quoted values for the group velocity were determined from the measured temporal delay of signal beam pulses passing through the rubidium cell. The quoted error bars here give the estimated uncertainty of the delayed pulse peak position obtained by fitting a Gaussian curve to the measured temporal intensity distribution.

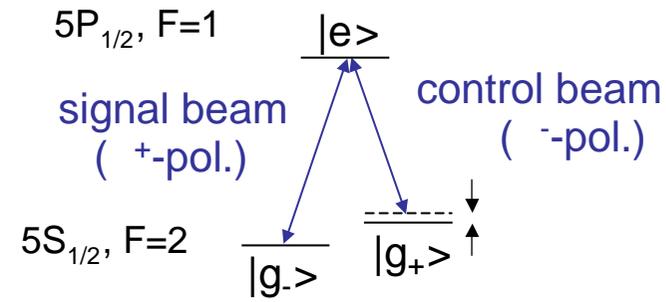

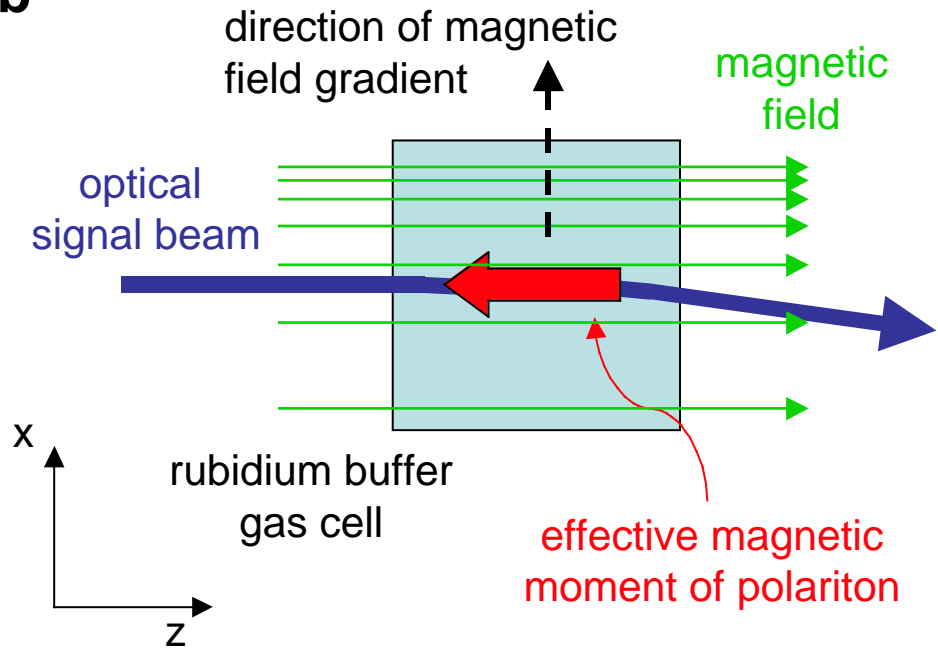

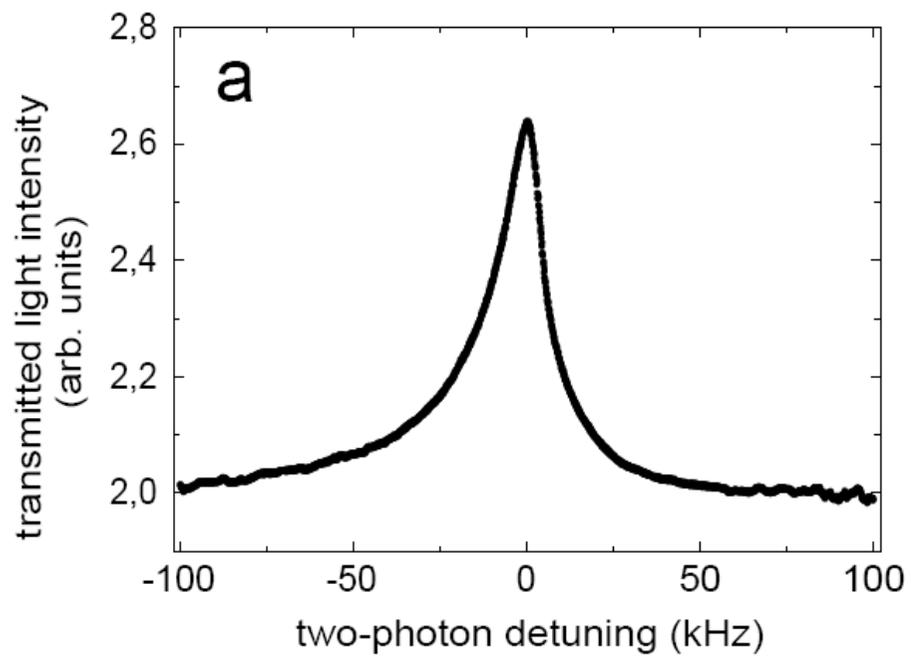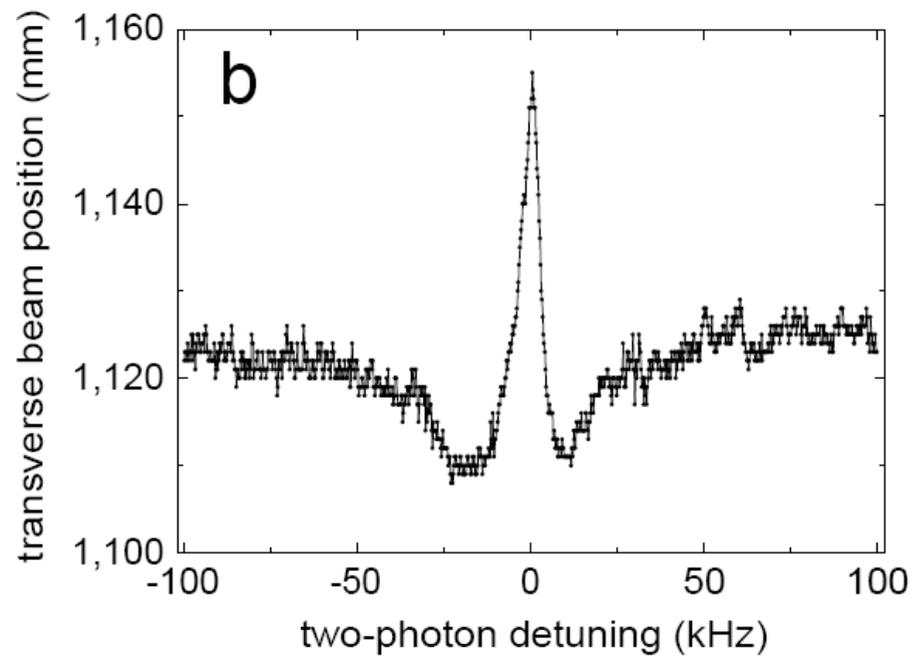

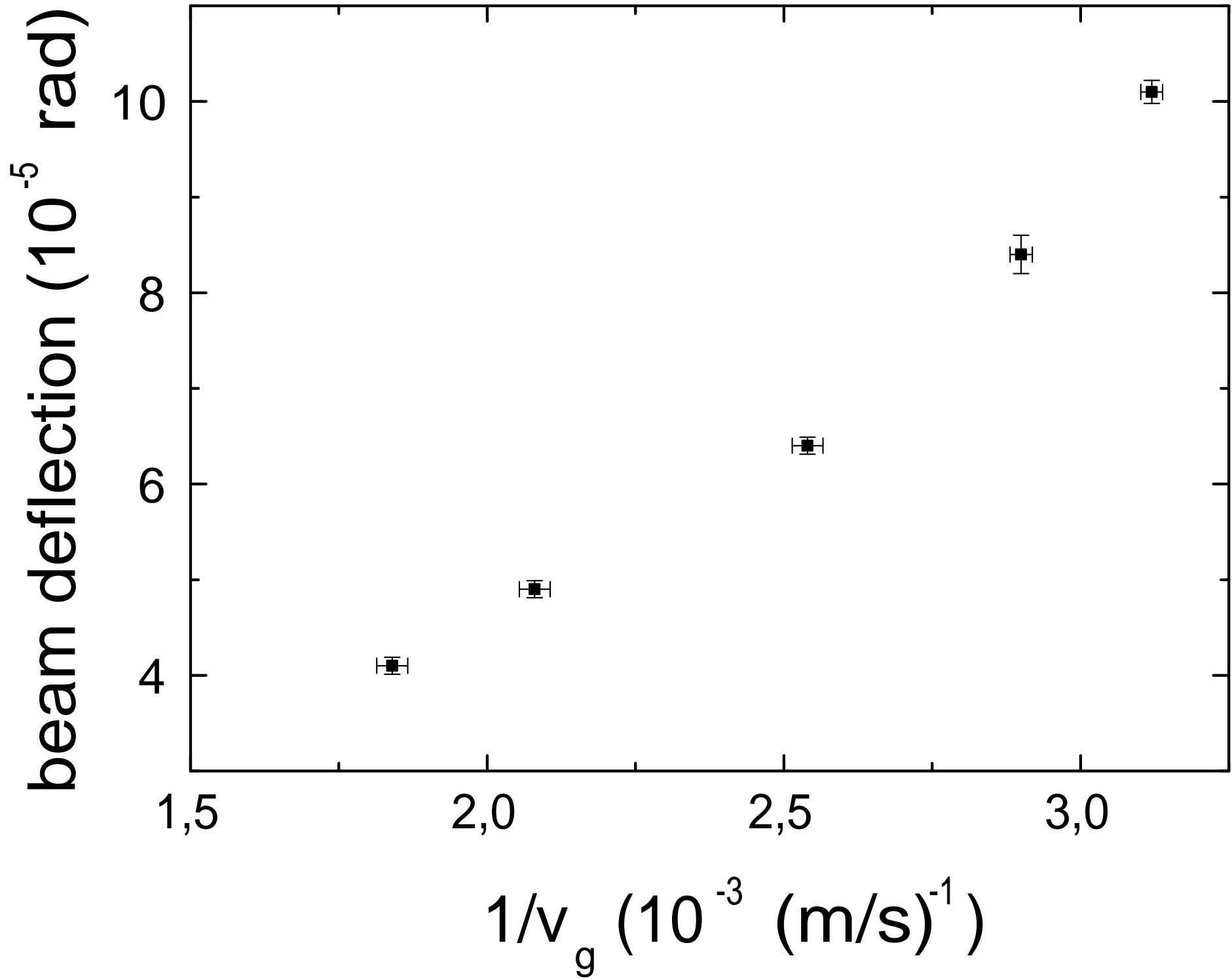